\def\et{$E_T$}                          
\def\met{\mbox{${\hbox{$E$\kern-0.6em\lower-.1ex\hbox{/}}}_T$}} 
\def\D0{D\O}                            
\def\ppbar{$p\overline{p} $}            

\def\etaj2lt{$|\eta_2| <$}
\def\gtetaj2lt{$< |\eta_2| <$}
\def\wjet{$W+${\rm jets}}
\def\rsig{$R_{\mathrm{sig}}$}
\documentstyle[preprint,epsfig,psfig,aps,floats,subfigure]{revtex}
\begin{document}

\tightenlines
\title{
Evidence of Color Coherence Effects in \wjet\ Events
from \ppbar\ Collisions at $\sqrt{s} = 1.8$ TeV }
%
%
\author{                                                                      
B.~Abbott,$^{45}$                                                             
M.~Abolins,$^{42}$                                                            
V.~Abramov,$^{18}$                                                            
B.S.~Acharya,$^{11}$                                                          
I.~Adam,$^{44}$                                                               
D.L.~Adams,$^{54}$                                                            
M.~Adams,$^{28}$                                                              
S.~Ahn,$^{27}$                                                                
V.~Akimov,$^{16}$                                                             
G.A.~Alves,$^{2}$
% ...
\\                                                                            
\vskip 0.30cm                                                                 
\centerline{(D\O\ Collaboration)}                                             
\vskip 0.30cm                                                                 
}                                                                             
\address{                                                                     
\centerline{$^{1}$Universidad de Buenos Aires, Buenos Aires, Argentina}       
\centerline{$^{2}$LAFEX, Centro Brasileiro de Pesquisas F{\'\i}sicas,         
                  Rio de Janeiro, Brazil}                                     
\centerline{$^{3}$Universidade do Estado do Rio de Janeiro,                   
                  Rio de Janeiro, Brazil}                                     
\centerline{$^{4}$Institute of High Energy Physics, Beijing,                  
                  People's Republic of China}                                 
\centerline{$^{5}$Universidad de los Andes, Bogot\'{a}, Colombia}             
\centerline{$^{6}$Universidad San Francisco de Quito, Quito, Ecuador}         
\centerline{$^{7}$Institut des Sciences Nucl\'eaires, IN2P3-CNRS,             
                  Universite de Grenoble 1, Grenoble, France}                 
\centerline{$^{8}$DAPNIA/Service de Physique des Particules, CEA, Saclay,     
                  France}                                                     
\centerline{$^{9}$Panjab University, Chandigarh, India}                       
\centerline{$^{10}$Delhi University, Delhi, India}                            
\centerline{$^{11}$Tata Institute of Fundamental Research, Mumbai, India}     
\centerline{$^{12}$Kyungsung University, Pusan, Korea}                        
\centerline{$^{13}$Seoul National University, Seoul, Korea}                   
\centerline{$^{14}$CINVESTAV, Mexico City, Mexico}                            
\centerline{$^{15}$Institute of Nuclear Physics, Krak\'ow, Poland}            
\centerline{$^{16}$Institute for Theoretical and Experimental Physics,        
                   Moscow, Russia}                                            
\centerline{$^{17}$Moscow State University, Moscow, Russia}                   
\centerline{$^{18}$Institute for High Energy Physics, Protvino, Russia}       
\centerline{$^{19}$Lancaster University, Lancaster, United Kingdom}           
\centerline{$^{20}$University of Arizona, Tucson, Arizona 85721}              
\centerline{$^{21}$Lawrence Berkeley National Laboratory and University of    
                   California, Berkeley, California 94720}                    
\centerline{$^{22}$University of California, Davis, California 95616}         
\centerline{$^{23}$University of California, Irvine, California 92697}        
\centerline{$^{24}$University of California, Riverside, California 92521}     
\centerline{$^{25}$Florida State University, Tallahassee, Florida 32306}      
\centerline{$^{26}$University of Hawaii, Honolulu, Hawaii 96822}              
\centerline{$^{27}$Fermi National Accelerator Laboratory, Batavia,            
                   Illinois 60510}                                            
\centerline{$^{28}$University of Illinois at Chicago, Chicago,                
                   Illinois 60607}                                            
\centerline{$^{29}$Northern Illinois University, DeKalb, Illinois 60115}      
\centerline{$^{30}$Northwestern University, Evanston, Illinois 60208}         
\centerline{$^{31}$Indiana University, Bloomington, Indiana 47405}            
\centerline{$^{32}$University of Notre Dame, Notre Dame, Indiana 46556}       
\centerline{$^{33}$Purdue University, West Lafayette, Indiana 47907}          
\centerline{$^{34}$Iowa State University, Ames, Iowa 50011}                   
\centerline{$^{35}$University of Kansas, Lawrence, Kansas 66045}              
\centerline{$^{36}$Kansas State University, Manhattan, Kansas 66506}          
\centerline{$^{37}$Louisiana Tech University, Ruston, Louisiana 71272}        
\centerline{$^{38}$University of Maryland, College Park, Maryland 20742}      
\centerline{$^{39}$Boston University, Boston, Massachusetts 02215}            
\centerline{$^{40}$Northeastern University, Boston, Massachusetts 02115}      
\centerline{$^{41}$University of Michigan, Ann Arbor, Michigan 48109}         
\centerline{$^{42}$Michigan State University, East Lansing, Michigan 48824}   
\centerline{$^{43}$University of Nebraska, Lincoln, Nebraska 68588}           
\centerline{$^{44}$Columbia University, New York, New York 10027}             
\centerline{$^{45}$New York University, New York, New York 10003}             
\centerline{$^{46}$University of Rochester, Rochester, New York 14627}        
\centerline{$^{47}$State University of New York, Stony Brook,                 
                   New York 11794}                                            
\centerline{$^{48}$Brookhaven National Laboratory, Upton, New York 11973}     
\centerline{$^{49}$Langston University, Langston, Oklahoma 73050}             
\centerline{$^{50}$University of Oklahoma, Norman, Oklahoma 73019}            
\centerline{$^{51}$Brown University, Providence, Rhode Island 02912}          
\centerline{$^{52}$University of Texas, Arlington, Texas 76019}               
\centerline{$^{53}$Texas A\&M University, College Station, Texas 77843}       
\centerline{$^{54}$Rice University, Houston, Texas 77005}                     
}                                                                             
\date{\today}

%
\maketitle
\vspace{-.4cm}
\begin{abstract}
We report the results of a study of color coherence effects in
$p\overline{p}$ collisions based on data collected by the D\O\
detector during the 1994--1995 run of the Fermilab Tevatron Collider,
at a center of mass energy $\sqrt{s} = 1.8$~TeV.  Initial-to-final state color 
interference
effects are studied by examining particle distribution patterns in
events  with a $W$ boson and at least one jet. The data are 
compared to Monte Carlo simulations with
different color coherence implementations and to an analytic
modified-leading-logarithm perturbative calculation based on the local
parton-hadron duality hypothesis.   
\end{abstract}


\newpage

    Color coherence phenomena in the final state have been studied since the
early 1980's in $e^+e^-$ annihilations\cite{cc1,cc2,cc3,cc4,cc5,cc5a} and are 
very well established.
The study of coherence effects in hadron--hadron collisions is considerably
more subtle than those in $e^+e^-$ annihilations due to the presence of colored
constituents in both the initial and final states.  
In this paper
we report the first results on initial-to-final state color interference 
effects in \ppbar\ interactions using \wjet\ events.

Coherence phenomena are an intrinsic property of any gauge theory.  
In quantum chromodynamics (QCD), color coherence phenomena can be 
instructively separated into two regions: 
intrajet and interjet coherence\cite{ccgen1,ccgen2}.  
Intrajet coherence deals with coherent effects in partonic cascades,
resulting on average in the angular ordering (AO) of the sequential parton
branches, which give rise to the 
depletion of soft particle production (the so called ``hump-backed plateau") 
inside jets\cite{lphd,dla1,intra1}.  
Interjet coherence is responsible for the string\cite{cc6} or 
drag\cite{cc7} effect first observed in the final state products of 
$e^+e^-$ annihilations.  It
deals with the angular structure of soft particle flows when three or more
energetic partons are involved in the hard process.  
In this case, the overall structure of particle angular distributions 
is governed by the underlying color dynamics of the hard scattering processes
at short distances. 

Perturbative quantum chromodynamics (pQCD) calculations have been used to 
describe the production of jet final states.  However, descriptions of the 
characteristic particle structure of high energy hard collisions 
still rely on phenomenological models to explain how the partonic cascade 
evolves into final state hadrons.
Within Monte Carlo (MC) simulations incorporating such models, the primary
partons from the hard scatter evolve into jets of 
partons via gluon and quark emission according to pQCD.
This process
continues until a cut-off transverse-momentum scale ($Q_0 \approx 1$~GeV$/$c) 
is reached.  After 
this phase, non-perturbative processes take over, which ``cluster" the final 
partons into color singlet hadronic states via a mechanism described by
phenomenological fragmentation
models, like the Lund ``string"\cite{string} or the ``cluster"\cite{cluster}
fragmentation models.
These simulations usually involve many {\it a priori} unknown 
parameters that need to be tuned to the data. 

A different and purely analytical approach giving quantitative predictions of
hadronic spectra is based on the concept of Local Parton Hadron Duality
(LPHD)\cite{lphd}.  The key assumption of this hypothesis is that the 
particle yield is described by a parton cascade in which the conversion of
partons into hadrons occurs at a low virtuality scale, of the order of hadronic
masses ($Q_0 \approx 200$~MeV/c$^2$) and independent of the scale of the 
primary hard 
process, and involves only low momentum transfers. It is assumed that the 
results obtained for partons apply to hadrons as well in an inclusive and
average sense.  Within the LPHD approach, resummed pQCD calculations for the 
parton cascade have been carried out in the
simplest case (high energy limit) in the Double Logarithmic Approximation 
(DLA)\cite{dla1,dla2},  and in the Modified Leading Logarithmic Approximation 
(MLLA)\cite{lphd,mlla1,mlla2}, which includes higher
order terms of relative order $\sqrt{\alpha_s}$ (e.g., finite energy 
corrections).  These higher order terms are essential for quantitative
agreement with data at present energies\cite{ccgen1,pic98}.  

    The AO approximation is an important consequence of color coherence.
It results in the suppression of soft gluon radiation in partonic cascades
in certain regions of phase space.  For the case of outgoing
partons, AO requires that the emission angles of soft gluons decrease
monotonically as the partonic cascade evolves away from the hard process. 
MC simulations including coherence effects probabilistically by means 
of AO are available for both initial and final state 
evolutions. (Parton shower event generators incorporate 
AO effects in the initial
state as the time reversed process of the outgoing partonic
cascade, i.e. the emission angles increase for the incoming partons as the
process develops from the initial hadrons to the hard subprocess.)
AO is an element of the DLA and MLLA analytic pQCD calculations,
which provides the probabilistic interpretation of soft-gluon cascades.  In
fact, beyond the MLLA a probabilistic picture of the parton cascade evolution 
is not
feasible due to $1/N_c^2$-suppressed (where $N_c$ is the number of colors) 
soft interference contributions that appear in the 
higher-order calculations\cite{ccgen2,mlla1}.

Both the CDF\cite{cc10} and D\O\cite{d0_coh} Collaborations have measured
spatial correlations between the softer third jet and the second leading-$E_T$
jet in $p\overline{p}~\rightarrow~3~\textit{jets}~+~X$ events
to explore the initial-to-final state coherence effects in \ppbar\
interactions.  The extraction of the color coherence signal in these
measurements relies on comparisons of data distributions to MC simulations with
and without coherence effects.
In the analysis described here, the 
coherence signal in the data is extracted in a more direct way by comparing 
the soft particle angular distributions around the colorless $W$ boson and 
opposing jet in the same event.

   The \D0\ detector is described in detail elsewhere \cite{D0detector}.  This
analysis uses the tracking system and the uranium/liquid-argon sampling 
calorimeter.  
The \D0\ calorimeter has a transverse granularity of
$\Delta\eta\times\Delta\phi = 0.1\times0.1$ forming projective towers, where 
$\eta$ is the pseudorapidity
($\eta=- \ln[\tan(\theta/2)]$, $\theta$ is the polar angle 
with respect to the proton beam), and $\phi$ is the azimuthal angle.
It has hermetic coverage for $|\eta|<4.1$ with fractional transverse energy
$E_T$ resolution of $\approx 80\%/\sqrt{E_T\textup{(GeV)}}$ for jets and 
fractional energy resolution of $\approx 15\%/\sqrt{E\textup{(GeV)}}$ for 
electrons.

The data sample for this analysis\cite{jaques}, representing an integrated
luminosity of 85~pb$^{-1}$, was collected during the 1994--1995 Tevatron 
Collider run.
Events from $W \rightarrow e+\nu$ decays were collected with a trigger that
required a minimum missing transverse energy (\met) of 15~GeV and an isolated
electromagnetic (EM) cluster with transverse energy $E_T > 20$~GeV.  The 
offline kinematic requirements imposed on this sample were 
$\met>25$~GeV, $E_{T}^{e}>25$~GeV, $|\eta_{e}| < 1.1 $, and
$40$~GeV/c$^2<M_T(e,\met)<110$~GeV/c$^2$, where 
$M_T(e,\met)=\sqrt{2p_T(e)p_T(\nu)(1-\textup{cos}\Delta\phi)/c^2}$ is the $W$ 
boson
transverse mass and $\Delta\phi$ is the azimuthal separation between the
electron and neutrino.  The transverse momentum of the neutrino, $p_T(\nu)$, was
calculated using the calorimetric measurement of the \met\ in the event.

Events were required to have one electron cluster passing 
four quality criteria based on shower profile and
tracking information: (i) the ratio of the EM energy to the total shower energy
had to be greater than 0.95, (ii) the position of the calorimeter energy 
deposition of
the electron had to match with a track found in the drift chambers,
(iii) the lateral and longitudinal shape of the energy cluster
had to be consistent with those of an electron, and 
(iv) the electron had to be isolated from other energy
deposits in the calorimeter with isolation fraction 
$f_{\mathrm{iso}} < 0.1$.
The isolation fraction is defined as 
$f_{\mathrm{iso}}=[E(0.4)-E_{\mathrm{EM}}(0.2)]/E_{\mathrm{EM}}(0.2)$, where 
$E(R_{\mathrm{cone}})$ ($E_{\mathrm{EM}}(R_{\mathrm{cone}})$) is the total 
(electromagnetic) energy within a cone of
radius $R_{\mathrm{cone}}=\sqrt{(\Delta\eta)^2 + (\Delta\phi)^2}$
centered around the electron.  

    Jets in the events were reconstructed offline using an iterative 
fixed-cone clustering algorithm with cone radius 
$R_{\mathrm{cone}}=0.7$.
Spurious jets from isolated noisy calorimeter cells and accelerator losses
were eliminated by loose cuts on the jet shape.  

Events were required to have a measured vertex with longitudinal position 
within 20~cm of the detector center to preserve the projective geometry.  
Since multiple interactions (more than
one proton--antiproton interaction in the same bunch crossing) are expected to
increase the global energy level in the calorimeter affecting the 
color coherence signal, we retained only events with a
single reconstructed vertex and additionally required the beam-beam hodoscope
timing information to be consistent with a single interaction.
Finally, events were eliminated when there was significant pileup 
energy in the calorimeter around the region where the 
Tevatron Main Ring passes through the \D0\ detector.

We study color coherence in \wjet\ events by comparing the
distributions of soft particles around the $W$ boson and opposing
leading-$E_T$ jet. Since the $W$ boson is a colorless object, it 
should not contribute
to the production of secondary particles, thereby providing a template
against which the pattern around the jet may be compared. This
comparison reduces the sensitivity to global detector and underlying
event biases that may be present in the vicinity of both the $W$ boson
and the jet.

The $W$ boson was
reconstructed from its electron and neutrino decay products, resulting in a 
twofold ambiguity in the $W$ boson rapidity $y_W$ (due to the corresponding 
ambiguity in the neutrino longitudinal momentum $p_Z$). MC
studies have shown that the smaller $|y_W|$ is closer to the true $W$ boson
rapidity approximately
2/3 of the time, so this is the solution chosen. This choice was also
made in the MC $W$ boson reconstruction to retain consistency
in the comparison with the data.

\begin{figure}[htb]
\centerline{\psfig{figure=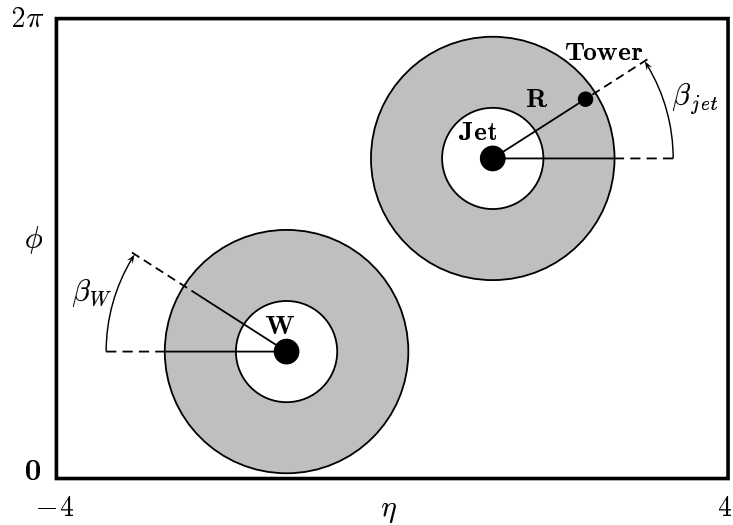,width=11cm}}
 \caption{Annular regions around the $W$ boson and the tagged jet 
          in $(\eta,\phi)$ space.}
 \label{PL_WJET_FIG1}
 \vspace{0.1cm}
\end{figure}

Once the $W$ boson direction has been determined, the
opposing jet was identified by selecting the leading-$E_T$ jet in the
azimuthal hemisphere opposite to the $W$ boson.  Annular regions are drawn
around both the $W$ boson and the tagged jet in $(\eta,\phi)$ space.
The angular distributions of 
calorimeter towers with \et $>$ 250~MeV are
measured in these annular regions using the polar variables
$R=\sqrt{(\Delta\eta)^2 + (\Delta\phi)^2}$ and $\beta_{X} =
\tan^{-1}(\frac{sign(\eta_{X})\cdot\Delta\phi_{X}}
{\Delta\eta_{X}})$; where $X = W$ or jet, 
$\Delta\eta_{W} = \eta_{\mathrm{tower}} - y_W$,
$\Delta\eta_{\mathrm{jet}} = \eta_{\mathrm{tower}} - \eta_{\mathrm{jet}}$, and 
$\Delta\phi_{X} = \phi_{\mathrm{tower}} - \phi_{X}$, in a
search disk of $0.7 < R < 1.5$ (Fig.\ \ref{PL_WJET_FIG1}).  
For events in which the jet and the $W$ boson annuli overlapped, all 
calorimeter towers in the shared region were assigned to the nearest object.

We define $\beta_{W\mathrm{(jet)}}=0$ to point along the beam direction 
nearest to the $W$ boson (jet).
Calorimeter towers which correspond to partially instrumented calorimeter 
regions (regions between the central and end-cap calorimeter cryostats with
pseudorapidities $1.1<|\eta_{\mathrm{tower}}|<1.4$) were not
included in the $\beta$ distributions for either the data or MC simulations. 
Calorimeter cells that belong to the electron cluster in a cone of 
$R=0.3$ from the centroid center were eliminated.
We study the interference effects in regions $|\eta_{\mathrm{jet}}|<0.7$ and
$|y_W|<0.7$, requiring the tagged jet to have $E_T > 10$~GeV and the $W$ boson
$p_T > 10$~GeV/c. After application of all selection criteria 390 events 
remain.

The measured angular distributions are compared to the predictions of  
{\small PYTHIA} 5.7 \cite{pythia} parton shower event generator 
with different levels of color coherence effects and to an analytic pQCD
calculation of Khoze and Stirling\cite{cc_khoze_stirling} based on MLLA and
LPHD.  The {\small PYTHIA} MC sample was processed through a full 
{\small {GEANT}}-based detector simulation \cite{geant}.  
To best model the calorimeter noise effects 
in the MC simulations, we overlaid noise contributions for each calorimeter cell
from the data.  The generated events
were subsequently processed using the same
criteria employed for analyzing the data.

{\small {PYTHIA}} incorporates initial
and final state color interference effects by means of the AO approximation of
the parton cascades.  After the perturbative phase it
employs the Lund string fragmentation (SF) model (or an independent
fragmentation IF model) as the phenomenological model
to describe the non-perturbative hadronization process.  The SF model has been
supported by the observations of color coherence phenomena
in $e^+e^-$ annihilations.
In {\small {PYTHIA}}, the AO constraint can be turned off.
When both AO and SF are implemented, {\small {PYTHIA}} accounts for color
coherence effects at both the perturbative and non-perturbative
levels.  Turning off AO removes the perturbative contribution, and
using IF eliminates the non-perturbative
component.  

\begin{figure}[htb]
\centerline{\psfig{figure=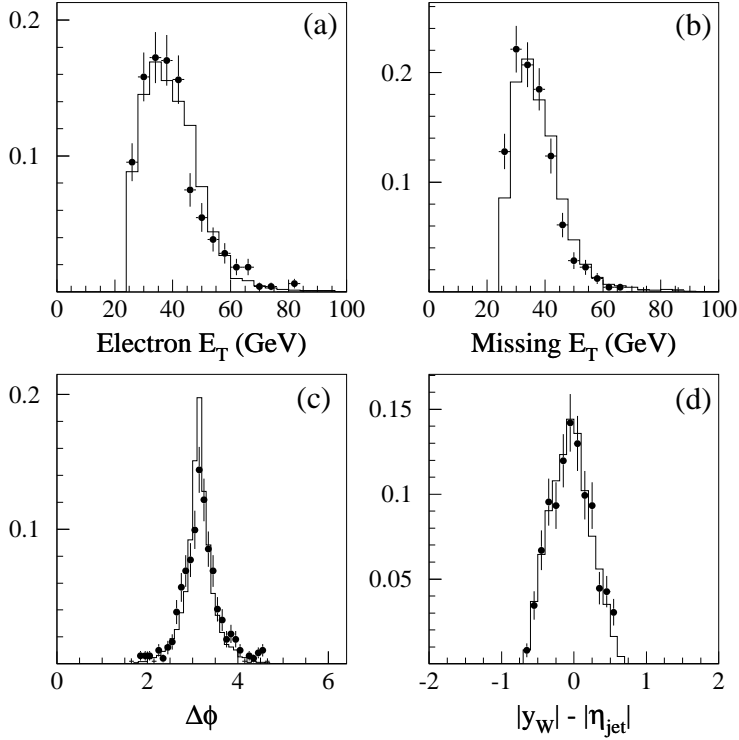,width=11cm}}
 \vspace{0.1cm}
    \caption{Comparisons of data (points) with {\small PYTHIA} events with AO
             and SF (histogram)
             for (a) electron \et, (b) event \met, (c) azimuthal separation of
             the $W$ boson and tagged jet, and (d) rapidity separation of the 
             $W$ boson and tagged jet.}
 \label{PL_WJET_FIG3}
 \vspace{0.1cm}
\end{figure}

We check whether the MC simulations describe the event characteristics in the
data using distributions of electron $E_T$, the event \met, and the
azimuthal and rapidity separation of the $W$ boson and the tagged jet.  These
are compared to the {\small PYTHIA} simulations with full coherence effects in
Fig.~\ref{PL_WJET_FIG3}.
For all distributions, {\small PYTHIA} is in good agreement with the data.

\begin{figure}[tb]
 \psfig{figure=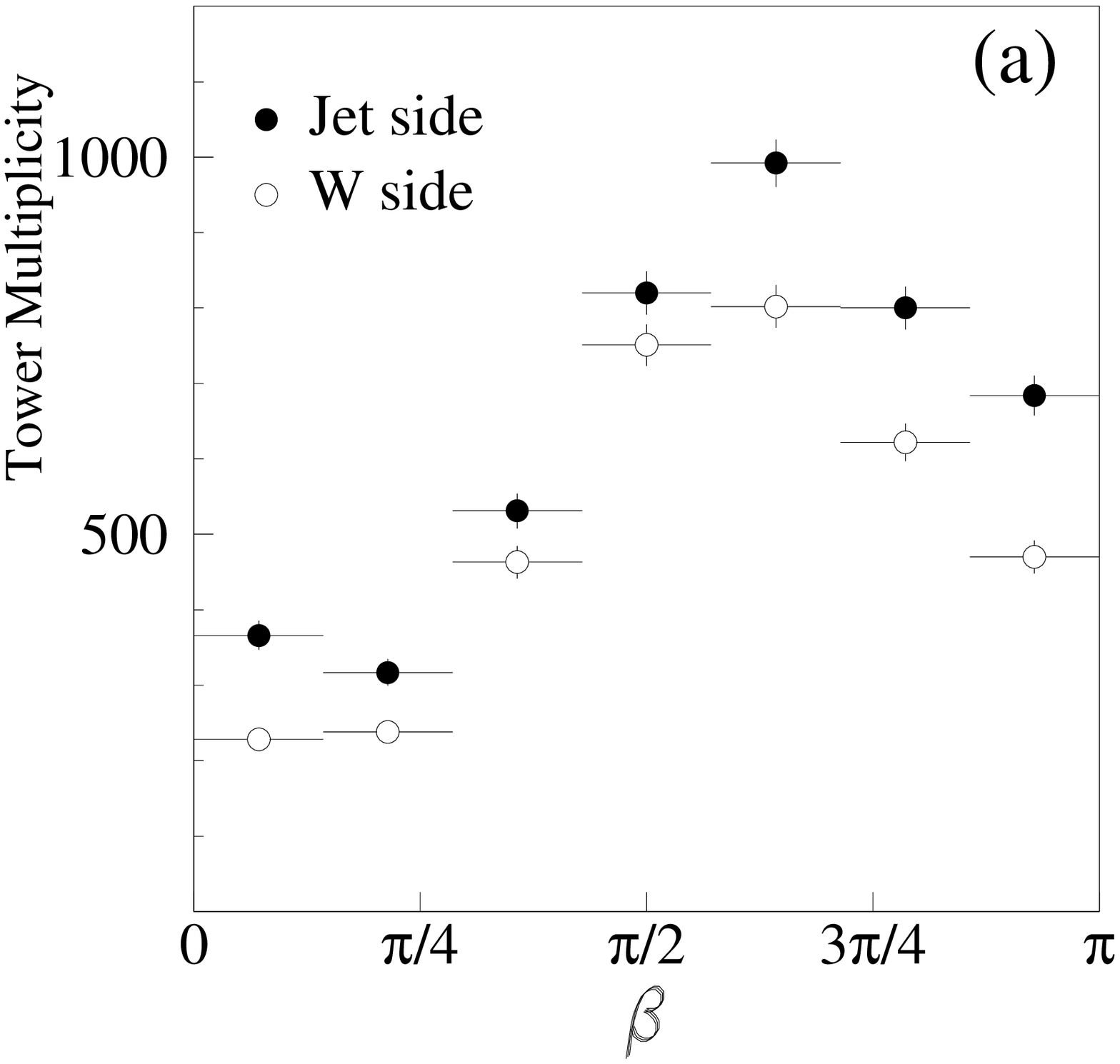,width=7.5cm,height=7.5cm}

  \vspace*{-7.5cm}

  \hspace*{8.1cm}
 \psfig{figure=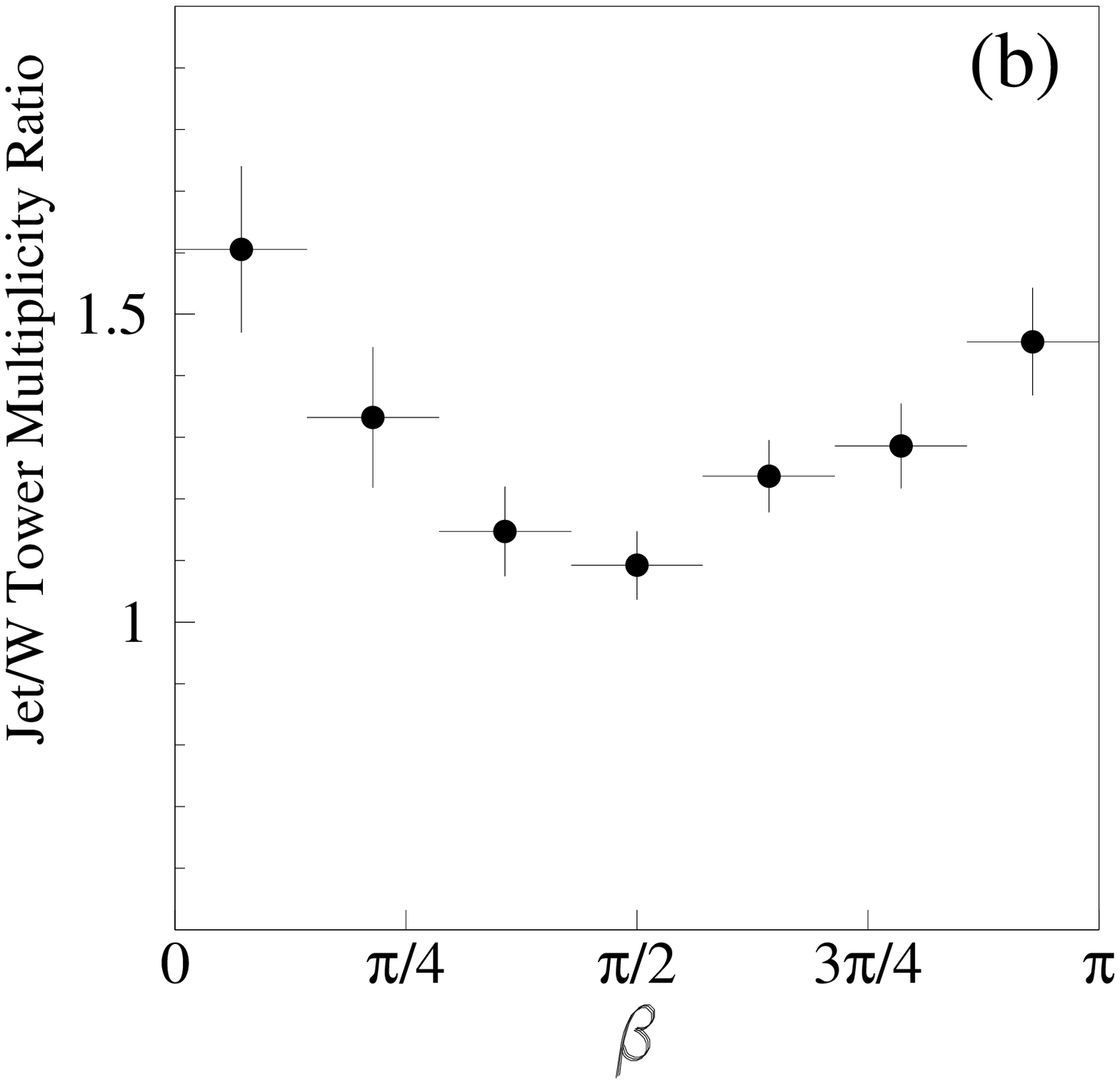,width=7.5cm,height=7.5cm}
 \vspace{0.1cm}
    \caption[]{(a) Calorimeter tower multiplicity around the jet and $W$ boson 
    as a function of $\beta$. 
    (b) Ratio of the tower multiplicity around the jet to the tower multiplicity
    around the $W$ boson as a function of $\beta$.}
 \label{PL_WJET_FIG2}
\end{figure}

Figure~\ref{PL_WJET_FIG2}(a) shows the measured angular distributions
of the number of towers above threshold around the jet and around the $W$ boson.
The $E_T$ of each calorimeter cell (each calorimeter tower is constructed from
many calorimeter cells following the projective geometry of the detector with
respect to the detector center)
was corrected for offsets due to noise, zero
suppression, and energy pileup effects\cite{jaques}. 
A prominent feature of both curves is a strong peaking around 
$\pi/2<\beta<3\pi/4$.  This is because the shapes of the $\beta$ distributions 
are sensitive both to process dynamics and to phase space effects resulting 
from our event selection criteria and the calorimeter tower $E_T$ threshold. 

Figure~\ref{PL_WJET_FIG2}(b) shows the ratio of the tower multiplicity
around the jet to the tower multiplicity around the $W$ boson
as a function of $\beta$. 
The errors include only statistical uncertainties, which are the dominant
source of uncertainty.
The data show that the tower multiplicity around the jet 
is greater than that around the $W$ boson and the excess is
enhanced in the event plane (i.e., the plane defined
by the directions of the $W$ boson or jet and
the beam axis: $\beta=0,~\pi$) and minimized in the
transverse plane ($\beta = \pi/2$).  
This is consistent with the
expectation from initial-to-final state color
interference that there is an enhancement of soft particle production
around the tagged jet in the event plane 
relative to the transverse plane when compared with the particle production 
around the $W$ boson.  It is also in agreement with our published 
multijet analysis results\cite{d0_coh}.

\begin{figure}[htb]
\centerline{\psfig{figure=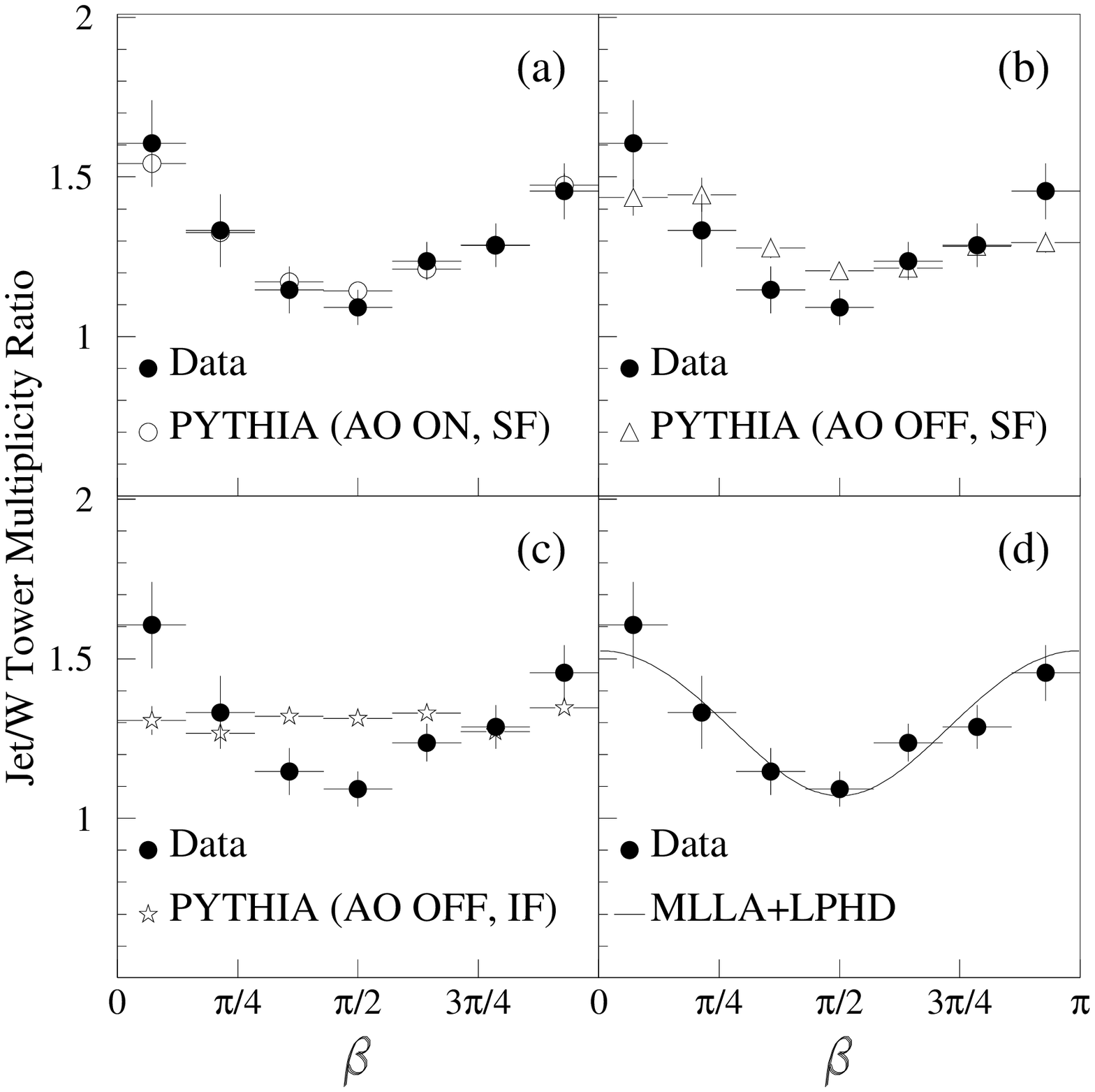,width=11cm}}
 \vspace{0.3cm}
    \caption{Comparison of the Jet/$W$ boson tower multiplicity ratio from data 
     to (a) {\small PYTHIA} with AO on and SF, (b) AO off and SF, (c) AO off
     and IF, and to the (d) MLLA+LPHD predictions.  The predictions have been
     normalized to the data.  All uncertainties are statistical only.}
 \label{PL_WJET_FIG4}
 \vspace{0.1cm}
\end{figure}

The ratio of the tower multiplicity
around the jet to the tower multiplicity around the $W$ boson
is shown in Fig~\ref{PL_WJET_FIG4} for the data and the MC predictions 
as a function of $\beta$.  All predictions have been
normalized to the integral of the data $\beta$-distribution.  {\small PYTHIA}
with AO and SF is in good agreement with the \wjet\
data. {\small PYTHIA} with AO off and SF agrees less
well, and {\small PYTHIA} with
AO off and IF does not reproduce the data.  
The shape of the analytic prediction based on MLLA and LPHD is 
consistent with
the data as shown in Fig.~\ref{PL_WJET_FIG4}(d).  We note that 
the analytic calculation was performed assuming that the $W$ boson and 
the outgoing parton were produced centrally and does not include any underlying
event or detector simulation effects.  

\begin{figure}[tb]
\centerline{\psfig{figure=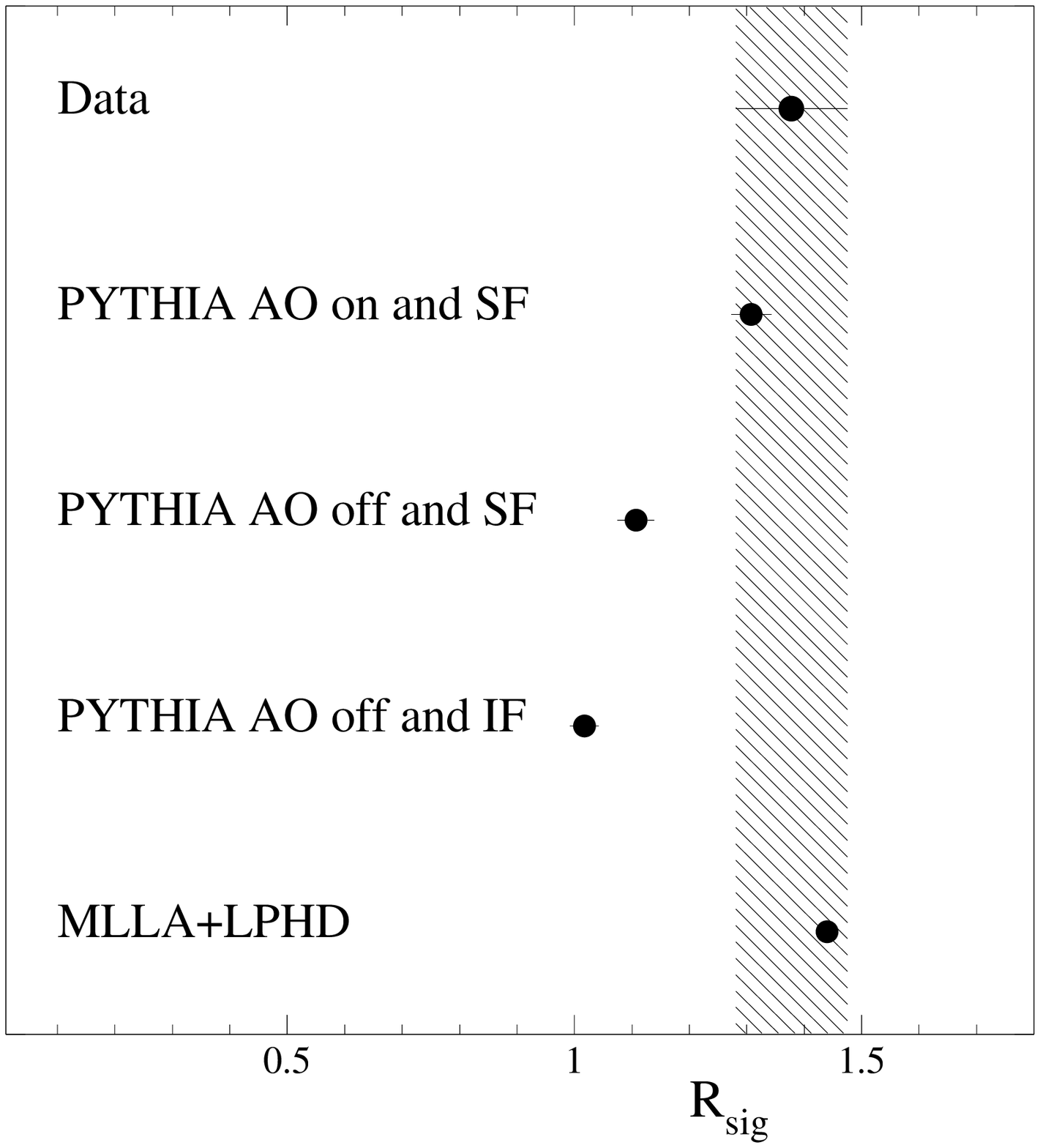,width=10cm}}
 \vspace{.2cm}
    \caption{\rsig\ for \D0\ data, {\small PYTHIA} with
          various coherence implementations, and a MLLA+LPHD QCD calculation.
          The errors are statistical only and the shaded band shows the
          statistical uncertainty on the \rsig\ variable for the data.}
 \label{PL_WJET_FIG5}
 \vspace{0.1cm}
\end{figure}

To measure the color coherence signal, we construct the variable \rsig\ as
the jet/$W$ boson tower multiplicity ratio of the event
plane ($\beta = 0,\pi$) to the transverse plane ($\beta = \pi/2$).  We define
$R_{\mathrm{sig}}=\frac{R(\beta = 0,\pi)}{R(\beta = \pi/2)}$, where 
$R(\beta = 0,\pi)=%
\frac{N_{\mathrm{jet}}^{1^{st} bin}+N_{\mathrm{jet}}^{7^{th} bin}}%
{N_W^{1^{st} bin}+N_W^{7^{th} bin}}$, 
$R(\beta = \pi/2)=\frac{N_{\mathrm{jet}}^{4^{th} bin}}{N_W^{4^{th} bin}}$, and
$N_{W\mathrm{(jet)}}^{i^{th} bin}$ is the number of towers above threshold for 
the $i^{th}$ bin of the $W$ boson (jet) $\beta$ distribution.
\rsig\ is expected to be near unity in the absence of color 
coherence effects.
In addition \rsig\ is insensitive to the overall normalization of
the individual distributions, and MC studies have shown that
it is relatively insensitive to detector smearing effects.
Figure~\ref{PL_WJET_FIG5} compares the \rsig\ variable for the 
data to the 
various {\small PYTHIA} predictions and to the MLLA+LPHD calculation.  
Clearly the value of \rsig\ for the data deviates from unity in agreement 
with {\small PYTHIA} with AO on and SF, and in disagreement
with AO off and SF or AO off and IF.  These comparisons imply 
that for the process under study, string fragmentation alone cannot 
describe the effects seen in the data.  The AO approximation is an element 
of parton-shower event generators that needs to be included if color 
coherence effects are to be modeled successfully.
Finally, the analytic prediction by Khoze and Stirling is consistent with 
the data, thus giving additional evidence supporting the 
validity of the LPHD hypothesis.

The dominant source of uncertainty on the data $\beta$ distributions is
statistical due to the limited event sample.  The statistical error for
\rsig\ is 7\%. 
Since we report ratios of event distributions any possible uncertainty on 
quantities 
that affect the overall rate of events is minimized.
Sources of systematic uncertainty arise from background contamination to 
\wjet\ events, uncertainties in the calorimeter channel-to-channel offset 
correction, multiple 
\ppbar\ interactions, and uncertainties associated with the calorimeter tower 
$E_T$ threshold.
The primary background to \wjet\ events is dijet production in which one of 
the jets mimics the characteristics of an electron. The \met\ in
such events typically arises from shower fluctuations or calorimeter
imperfections.  For our selection criteria, the estimated background level
is about $5$\% resulting in a 1-2\% uncertainty on the coherence signal.  
The effect on the \rsig\ variable due to uncertainties in the
offset correction were found to be at the 1\% level.  To evaluate possible 
effects on the signal coming from residual multiple 
interaction contamination, we examined the dependence of
\rsig\ as a function of luminosity.  No systematic dependence of the signal
variable was observed for our event sample.

\begin{figure}[htb]
\centerline{\psfig{figure=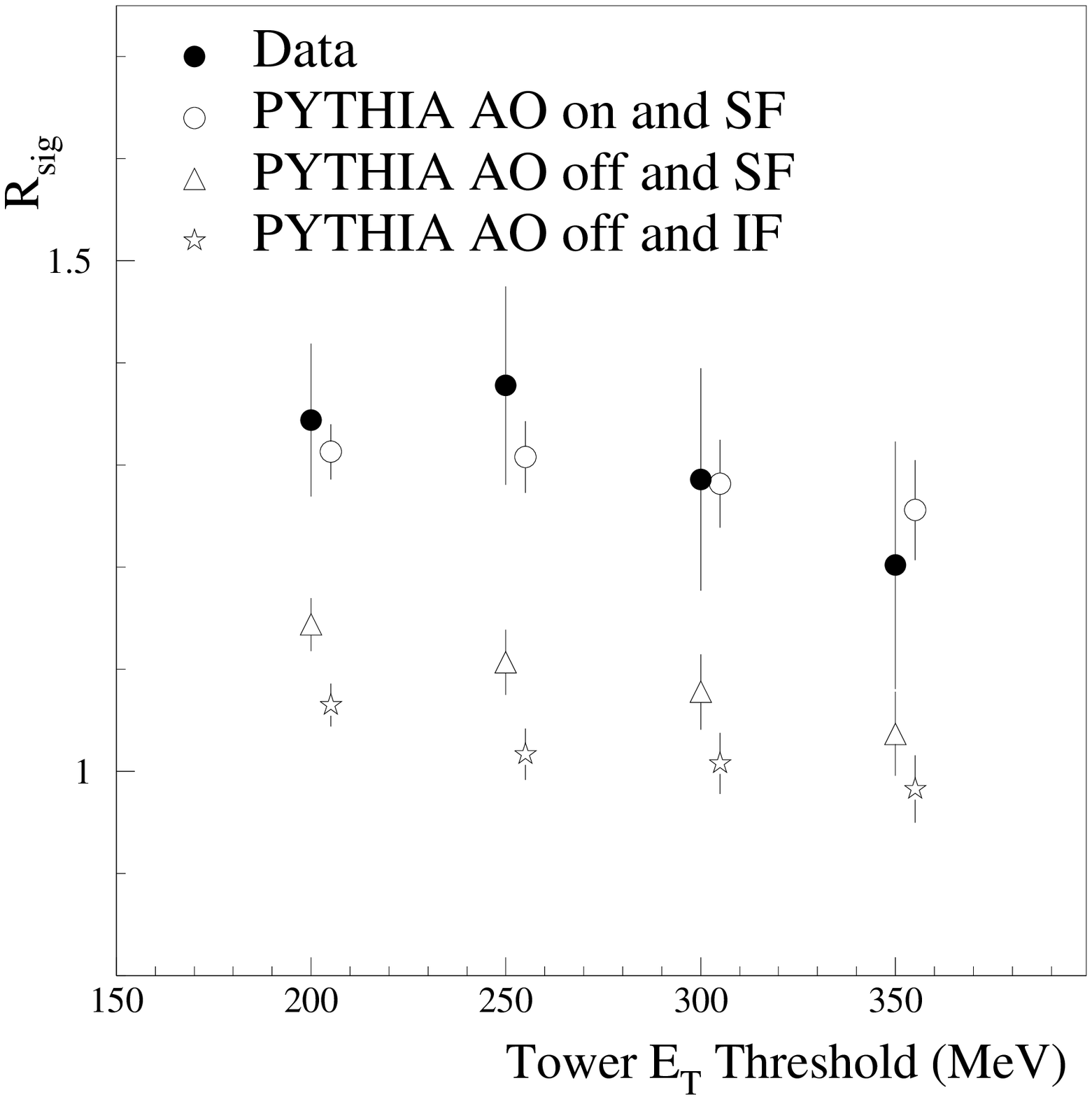,width=10cm}}
 \vspace{0.4cm}
    \caption{\rsig\ as a function of the tower $E_T$
        threshold for the data (solid circles) and for the various 
        {\small PYTHIA} predictions.  All uncertainties are statistical only.}
 \label{PL_WJET_FIG6}
 \vspace{0.1cm}
\end{figure}

The dependence of the signal variable \rsig\ on the calorimeter tower $E_T$
threshold was studied by varying the threshold from 200 to 350 MeV.
Figure~\ref{PL_WJET_FIG6} shows how the \rsig\ variable varies
as a function of the tower threshold for the data
and for the three {\small PYTHIA} predictions.  Although there seems to be a
dependence of \rsig\ with the tower $E_T$ threshold for both the data and the
simulations, for all tower thresholds
examined the {\small PYTHIA} predictions with AO on and with SF resemble the
data best.

In summary, color coherence effects in \ppbar\ interactions have been observed 
and studied by the \D0\ Collaboration.  We have presented the first
results on color coherence effects in \wjet\ events. The data
show an enhancement of soft particle radiation around the jet in the event 
plane with
respect to the transverse plane, consistent with color coherence as
implemented in the {\small PYTHIA} parton shower event generator, which
includes the angular ordering approximation and string fragmentation.  In 
addition, the relative amount of enhancement is consistent with an
analytic perturbative QCD calculation based on modified leading logarithmic
approximation and local parton-hadron duality.

We appreciate fruitful discussions with 
V. Khoze and J. Stirling.  We also thank T. Sj\"{o}strand
for helping us with the color coherence implementations in {\small {PYTHIA}}.
%
%
We thank the Fermilab and collaborating institution staffs for 
contributions to this work, and acknowledge support from the 
Department of Energy and National Science Foundation (USA),  
Commissariat  \` a L'Energie Atomique (France), 
Ministry for Science and Technology and Ministry for Atomic 
   Energy (Russia),
CAPES and CNPq (Brazil),
Departments of Atomic Energy and Science and Education (India),
Colciencias (Colombia),
CONACyT (Mexico),
Ministry of Education and KOSEF (Korea),
and CONICET and UBACyT (Argentina).

\end{document}